\documentclass[10pt]{article}

\usepackage{amsmath}
\usepackage{amssymb}

\usepackage{cite}

\usepackage{lineno}

\usepackage{microtype}

\usepackage[labelfont=bf,labelsep=period,justification=raggedright]{caption}
\RequirePackage{graphicx}

\makeatletter
\renewcommand{\@biblabel}[1]{\quad#1.}
\makeatother

\date{}

\begin{document}

\begin{flushleft}
{\Large
\textbf{Effect of noise in intelligent cellular decision making}
}
\\
R. Bates$^{1}$, 
O. Blyuss$^{2}$, 
A. Alsaedi$^{3}$, 
A. Zaikin$^{1,2,4,\ast}$
\\
\bf{1} Department of Mathematics, University College London, UK
\\
\bf{2} Institute for Women's Health, University College London, UK
\\
\bf{3} Department of Mathematics, King AbdulAziz University, Jeddah, Saudi Arabia
\\
\bf{4} Lobachevsky State University of Nizhniy Novgorod, Nizhniy Novgorod, Russia 
\\
$\ast$ E-mail: Calexey.zaikin@ucl.ac.uk
\end{flushleft}

\section*{Abstract}
Similar  to intelligent multicellular neural networks controlling human brains, even single cells surprisingly are able to make intelligent decisions to classify several external stimuli or to associate them. This happens because of the fact that gene regulatory networks can perform as perceptrons, simple intelligent schemes known from studies on Artificial Intelligence. We study the role of genetic noise in intelligent decision making at the genetic level and show that noise can play a constructive role helping cells to make a proper decision. We show this using the example of a simple genetic classifier able to classify two external stimuli.


\section*{Introduction}
Cellular decision making and the determination of cellular fate is performed by superimposing the pattern of extracellular signalling with the intracellular state given by locations and concentrations of chemicals within the cell. This process is regulated by the program encoded in the cellular genome. Understanding the principles of this programming and, especially, how this program is executed is a key problem of modern biology, for control of this program would provide insight into new biotechnologies and medical treatments. 


Our brains is able to make intelligent decisions because they are controlled by neural networks, i.e., networks based on the communication between huge amount of cells. But can intelligent solutions be executed at the level of genome? First, we should define what we will understand under intelligence. For this we will use some basic schemes developed in the field of Artificial Intelligence. One of the basic schemes of intelligence has been suggested by Frank Rosenblatt, who has called such systems as "perceptrons". Basic perceptron was able to classify several external stimuli and provide binary output. Could such an intelligent decision making be observed in the performance of gene regulatory networks? It was long known that cells can adapt to and anticipate the stress but it remains not completely clear whether this is a result of intelligent learning or something else. For example, in 2008 Saigusa et al. \cite{saigusa_amoebae_2008} have shown that amoebae, a single cell organism, can anticipate periodic events.  Naturally, a fundamental question arises, can a genetic network behave intelligently in the sense that it will learn an association or classification of stimuli? Recent theoretical studies have shown that it is, in principle,  possible. It was shown that neural network can be built on the basis of chemical reactions, if a reaction mechanism has neuron-like properties \cite{1991_Hjelmfelt}. In these works linked chains of chemical reactions could act as Turing machines or neural networks \cite{hjelmfelt_implementation_1995}. D. Bray has demonstrated that a cellular receptor can be considered as a perceptron with weights which  have been learned via genetic evolution \cite{1994_Bray}, showing formally that protein molecules may work as computational elements in living cells [5]. Gandhi et al. has formally shown that also associative learning can be performed in biomolecular networks \cite{2007_Gandhi}. In 2008 Fernando et al. have suggested a formal scheme of the single cell genetic circuit which can associatively learn within the cellular life \cite{fernando_molecular_2008}. The same team has investigated with positive result using the real genomic interconnections whether the genome of the bacterium E. Coli could work as a liquid state machine learning associatively how to respond to a wide range of environmental inputs \cite{2007_Jones}. Despite formal proof-of-the-principle experimental work has fallen short to fully implement genetic intelligence, e.g., in Synthetic Biology. To our knowledge, only L. Qian et al. have experimentally shown that neural network computations, in particular, a Hopfield associative memory, can be implemented with DNA gate architecture and DNA strand displacement cascades \cite{2011_Qian}. In their experiment, learning has been, however, executed in advance on the computer.

On the other hand, recently it has been demonstrated that gene expression is a very noisy process \cite{mcadams_its_1999}. 
Both intrinsic and extrinsic noise in a gene expression has been experimentally measured  in  \cite{elowitz_stochastic_2002} and modelled either with stochastic Langevin type
differential equations or with Gillespie-type algorithms to simulate the single chemical reactions underlying this stochasticity \cite{gillespie_chemical_2000}. Hence,  the question arises  as to what the fundamental role of noise in intracellular intelligence is. Can stochastic fluctuations only corrupt the information processing in the course of decision making or can they can also help cells to make intelligent decisions? During last three decades it was shown that under certain conditions in nonlinear systems noise can counterintuitively  lead to ordering, e.g.,  in the effect of Stochastic Resonance (SR) \cite{1981_Benzi_jpa}, which has found many manifestations in biological systems, in particular to improve the hunting abilities of the paddlefish \cite{russell_use_1999}, to enhance human balance control \cite{priplata_vibrating_2003}, to help brain's visual processing \cite{2002_Mori_prl}, or to increase the speed of memory retrieval \cite{UsherF00}. Here we will show that, surprisingly, the correct amount of noise in genetic decision making can produce an improvement in performance in classification tasks, demonstrating Stochastic resonance in a genetic decision making (SRIDM).

To show this we have designed a simple genetic network able to classify two external stimuli.
The form in which our intelligence will take in this paper will be in the ability to perform linear classification tasks. Linear classification describes the ability to successfully discriminate between two sets which are linearly separable. Let $S_1$ and $S_2$ be linearly separable sets in $n$-dimensional space with a separation threshold $T$, i.e.,
\[
\exists \, w \in \mathbb{R}^n \hspace{0.7cm} \text{and} \hspace{0.7cm} \exists \, T \in \mathbb{R} 
\] 
Such that:
\[
\forall x_1 \in S_1, \hspace{0.4cm} w.x_1 < T \hspace{0.4cm} \text{and} \hspace{0.4cm} \forall x_2 \in S_2, \hspace{0.4cm} w.x_2 > T 
\]
Linear separation problems represent what could be described as one of the simplest
form of intelligence that a system could display. But although they may be simple they allow a system to
perform a large class of discrimination tasks. Many functions a cell may need to perform
require a yes or no decision to be made to proceed, the function of linear classifcation
allows a system to take in information from various sources (most likely chemical concentrations),
perform some analysis on this (take a weighted sum of the inputs) and make a final
decision based on this analysis, namely ``does this value exceed my required threshold
of activation?". There are countless examples of cells having to perform some kind of
quantitative `weighing up' of information from various sources such as in chemotaxis,
regulatory checkpoints (cyclin concentrations) and apoptosis or `cellular suicide' \cite{alberts_molecular_2008}. Additionally, if a separation threshold $T$ 
is linked to a current cellular state, such a scheme can learn the classification rule as an intelligent perceptron.

Linear classification algorithms of this nature are named perceptrons and are formulated in the following way, for $\forall x \in (S_1,S_2)$:

\begin{equation}
\textbf{x.w} =  \sum_{i=1}^{n} x_i. w_i  = \bar{x}
\label{sum1}
\end{equation}

\begin{equation}
  F(\bar{x}) = \left\{ 
  \begin{array}{l l}
    1 & \quad \text{if $\bar{x} > T$}\\
    0 & \quad \text{if $\bar{x} \leq T$}\\
  \end{array} \right.
\label{percep1}
\end{equation}
Here $
\textbf{x} = (x_1,x_2 \dots, x_n) 
$
represents the n-dimensional input and the weight vector $\textbf{w}= (w_1,w_2,\dots,w_n)$ and the value $T$ describe the separation hyperplane. As can clearly be seen from \eqref{sum1} and \eqref{percep1} the value of $T$ can be normalized to 1 by redefining,
 \[
\bar{w_i} = \frac{w_i}{T} \hspace{0.6cm} \forall i
\] 
From this we can see that our entire classification is now specified by the weight vector $\textbf{w}$. We also introduce the concept of a spoiled perceptron whereby the value of $T$ is adjusted slightly. This will clearly have the effect of causing the misclassification of points close to the separation hyperplane. 

\section*{Model of the linear genetic classifier}

The model we will consider is based on the designs of Kaneko \emph{et al.} \cite{suzuki_oscillatory_2011} for an arbitrarily connected $n$-node genetic circuit. Gene activity is modelled by differential equations coupled through Hill functions to describe inhibitory or activatory influence \cite{1997_Weiss}. We have utilized this framework to produce a gene regulatory circuit capable of linear separation on 2 nodal concentrations with the result exhibited as the steady state concentration reached on a third node, maximal concentration or zero concentration depending on the classification. The circuit diagram is as shown in Fig.\ref{fig_1}A with pointed arrowheads representing an promoting transcription factor and a flat-headed arrow representing a repressive transcription factor.

Letting $m(i,t)$ denote the $i$th nodal mRNA concentration and $p(i,t)$ denote the $i$th nodal protein concentration the rate equations are as follows:
\begin{equation}
\frac{dm(i,t)}{dt} = \gamma (F(i,t) - m(i,t))
\end{equation}
with
\begin{equation}
F(i,t) = f \left( \sum_j A_{ij}p(j,t) - \theta _i \right)
\end{equation}
\begin{equation}
f(x) = \frac{1}{1+e^{-\beta x}}
\end{equation}
and
\begin{equation}
\frac{dp(i,t)}{dt} = (m(i,t) - p(i,t))
\end{equation}

Where $\beta$ is some positive constant (taken in the following calculations to be $40$) and $f(x)$ is a sigmoid function switching from $0$ to $1$ if $x$ is increased. The matrix $A$ represents the topology of the genetic network. $A_{ij} > 0$ implies that the $i$th protein is an activating promoter for the $j$th gene, $A_{ij} < 0$ implies that the $i$th protein is an inhibitor for the $j$th gene and $A_{ij} = 0$ implies that there is no transcriptional relationship between the $i$th protein and $j$th gene. For example the $A$ matrix for the network in Fig.\ref{fig_1}a would be as follows:
\begin{equation}
A = \left( 
\begin{array}{ccccc}
0&0&0&-1&1\\
0&0&0&w_1&-w_1\\
0&0&0&w_2&-w_2\\
0&0&0&1&-1\\
0&0&0&-1&1
\end{array}
\right)
\label{Amatrix}
\end{equation}
It can clearly be seen that the dynamics of this system will be bounded, indeed we can easily show that if both $m(i,0)$ and $p(i,0)$ are in $[0,1]$ they will never leave this range. As a generalised chemical concentration could not be so rigourously contained we should consider these as concentrations relative to some maximal concentration level. 

As discussed earlier we wish to have some mechanism by which the weightings (as manifested by the values in the matrix \ref{Amatrix}) can be adjusted. As this would represent a multiplicational transcriptional strength we consider a mechanism of phosphorylation. We now require that the proteins produced at nodes 1 \& 2 must be activated via phosphorylation by some external kinase concentration. In this way we can consider the rate of production of phosphorylated protein as the product of concentrations of the unphosphorylated protein and the phosphorylating kinase concentration. It is by this mechanism that we justify the ability to adjust the values of $w_1$ and $w_2$ and thus the weightings for linear classification.

\section*{Results demonstrating effect of a stochastic resonance in the genetic classifier}

\begin{figure}[!h]
\includegraphics[width=2.5in]{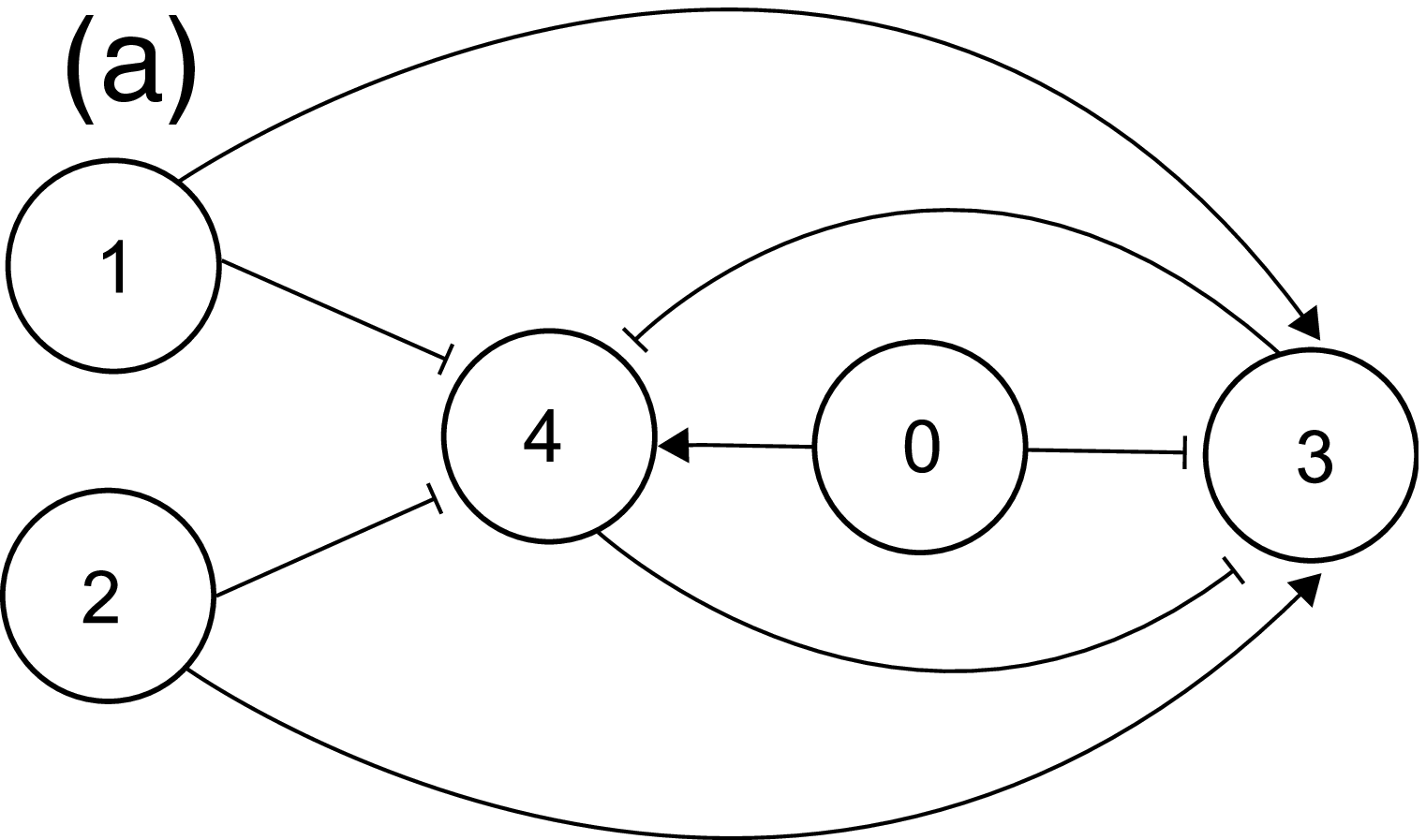}
\hspace{0.5cm}\includegraphics[width=2.0in]{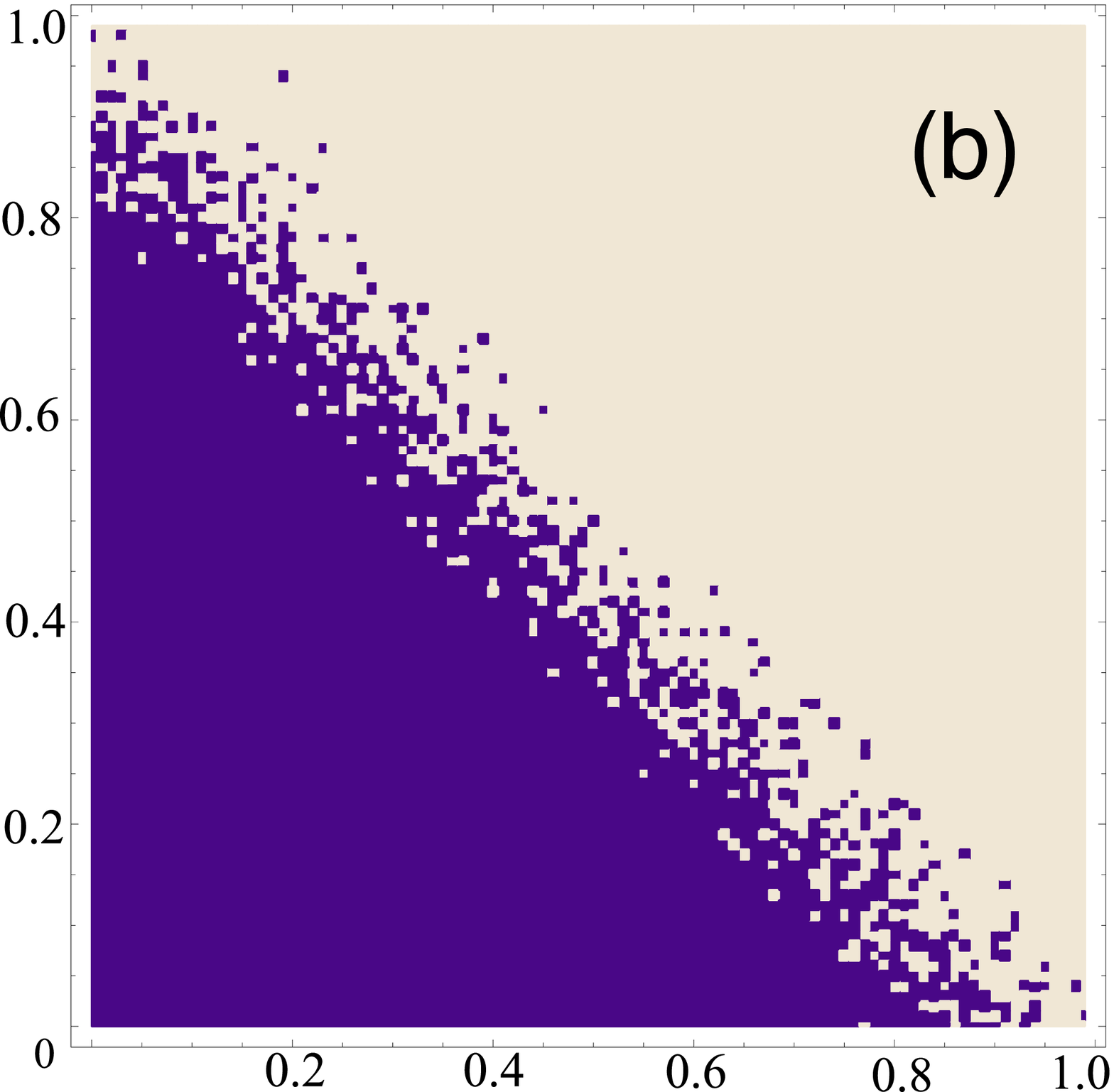}

\vspace{1.0cm}\centering\includegraphics[width=3.0in]{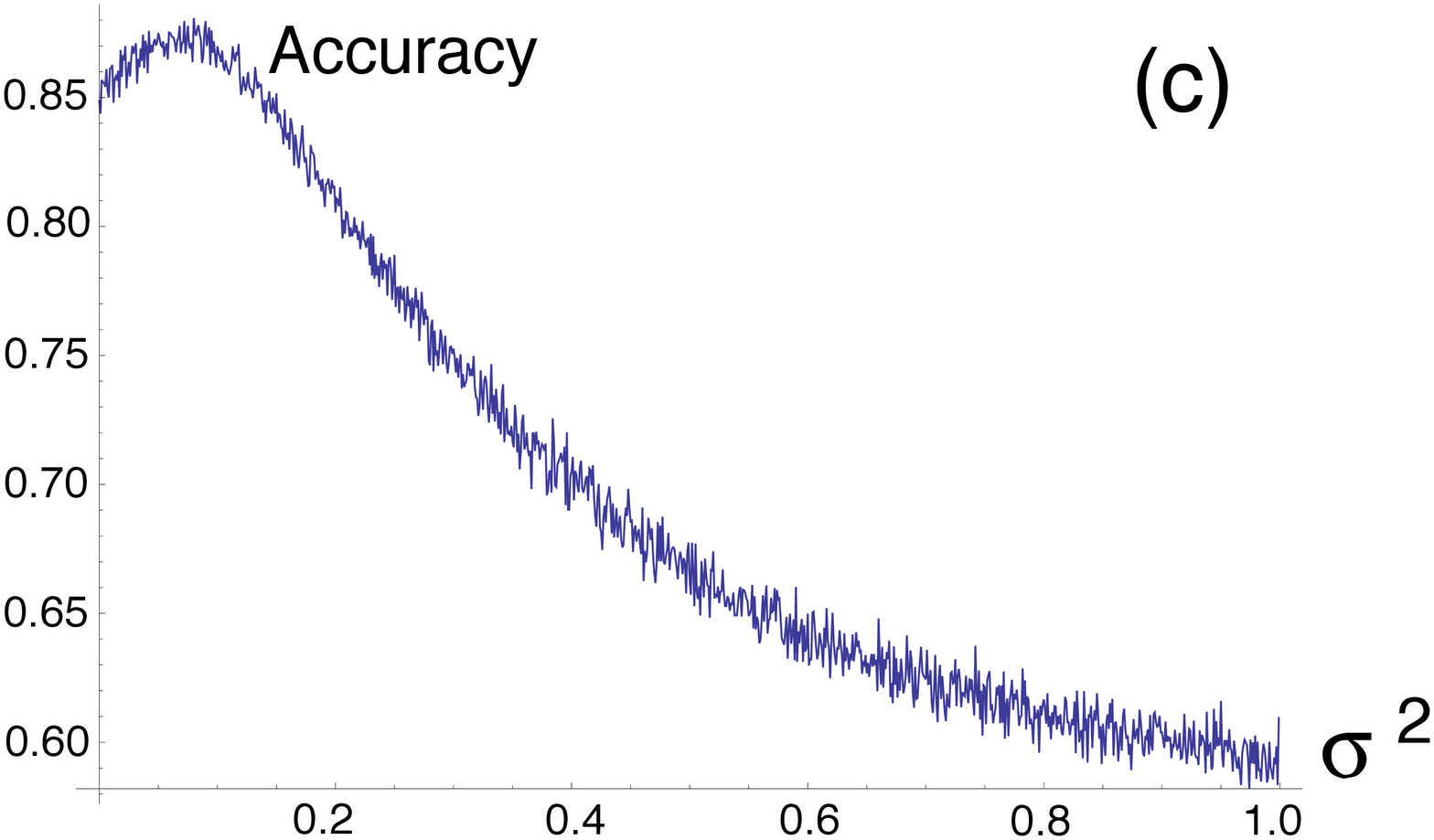}
\caption{(a) Design of an intracellular  gene regulatory network able to perform linear classification. In each cell the classifier is based on the toggle switch (genes 3 and 4, which, if separated, organize a bistable system in state ON-OFF or OFF-ON ) and  the gene 3  is the output. Two inputs are genes 1 and 2. Due to permanent basal expression of gene 0, the gene 4 is in the state ON. Correspondingly the output gene 3 is in the OFF state. If the join action of inputs 1 and 2 can repress gene 4 despite the activating link from 0, the switch will change its state and the output will be in the state ON. In this way the scheme can classify two inputs according to the binary classification. (b) The expression of gene 0 sets the ''threshold'' of classification. The slope of the separation line depends on the weights with which inputs 1 and 2 inhibit gene 4, activate gene 3, and on the expression of gene 0. The gene 0 can be one of the genes in the cellular genome, in this case, a classifier will ''learn'' the classification rule from surrounding. Here the example of linear separation for inputs 1 and 2 varying between 0 and 1 is shown, demonstrating the  effect of input noise in linear separation. Adding noise to inputs 1 and 2 blurs the classification line. (c) Stochastic resonance in a genetic perceptron. In a spoiled perceptron optimal amount of noise added to inputs improves the accuracy of a classification. 
}
\label{fig_1}
\end{figure}

In Fig.\ref{fig_1}b we demonstrate the systems ability to perform linear separation on the input initial conditions. In the current design the weightings by which the linear separation is performed is manifested in the strength of transcriptional regulation between certain nodes as seen in \eqref{Amatrix}. By varying this we are able to perform a wide range of linear classifcation tasks, i.e., we are able to perform any linear separation possible on the domain $[0,1]\times[0,1]$.
This adjustment of the transcriptional strength is justified earlier by the addition of a phosphorylation stage. 

From here onwards we will be interested in noisy systems where noise is an extrinsic noise added to inputs. Hence, rather than using techniques of stochastic simulation popular in Mathematical Biology we will be using deterministic systems adjusted by the addition of some stochastic term to inputs. Effect of this can clearly be seen in Fig.\ref{fig_1}b,  the addition of noise to the unspoiled threshold will only cause a decrease in accuracy of classification.
The situation will dramatically change if, initially, the classifier had some inaccuracy introduced by spoiling the threshold of classification. 
The context in which we will look for stochastic resonance is in the case that the mechanism has been spoiled in some sense. This will be manifested by an adjusted initial concentration value for the $0$th node. As stated earlier our systems performance without any spoiling of some kind is considered an ideal set of results so obviously no improvement can be made there and noise will only decrease our accuracy, this can be observed in Fig.\ref{fig_1}b  showing a severe decrease in accuracy with the addition of noise when there is zero spoiling.

For spoiled perceptron, however,  adding noise can surprisingly restore  a correct classification manifesting the effect of stochastic resonance.
The term stochastic resonance is somewhat of a misnomer as in this case we make no reference to acoustic or vibrational resonance. Instead what we mean is that some measure of functionality of our system is improved by the addition of some optimal intensity of noise. This optimal noise should be some finite value of the variance of an additive normally distributed noise term. This implies that our system should perform worse after a certain amount of noise.  This makes sense intuitively as an infinitely noisy system would not be able to facilitate any order.

The inputs to our system are generated as a pair of random numbers uniformly distributed between zero and one and then classified according to the unspoiled threshold. Next, the same inputs have the Gaussian noise added to them and are classified again this time using the spoiled threshold. The two classifications are then compared and compiled into the accuracy score as follows. Considering a set of $N$ input sets $(\bar{x}_1,\dots,\bar{x}_N)$ , let 
\[
O(i) , \hspace{1cm} (i=0,1,\dots,N)
\]  
be the non-noisy unspoiled value outputted for the $i$th input set, and let
\[
D(i) , \hspace{1cm} (i=0,1,\dots,N)
\]
be the noisy, spoiled, outputted value. 
Define 
\begin{equation}
T_+ = \{i:O(i)=D(i)=1\}
\end{equation}
and
\begin{equation}
T_- = \{i:O(i)=D(i)=0\}
\end{equation}
Then we can define accuracy as 
\begin{equation}
\label{acc}
\text{Accuracy} = \frac{|T_+| + |T_-|}{N}
\end{equation}

If we perform this accuracy analysis with a spoiled threshold we observe Fig.\ref{fig_1}c. Here we see that for some optimal value of input noise we witness an increase in accuracy.  This result is surprising as the noise added was normally distributed and thus directionally unbiased. This is also significant as we have genuinely produced an output set which is more faithful to the original than the non-noisy case. 

The governing equations for our genetic network model were simulated using a 4th order Runge-Kutta numerical integration scheme. Normally distributed random variables were generated using the Box-Muller transformation for uniform random variables.  When considering accuracy we calculated it in the following way, using a Monte Carlo style approach. We generated a set of 10000 pairs of random variables, distributed randomly on the interval $[0,1]$. These pairs became the initial concentrations of nodes 1 and 2 respectively. The network is then simulated for these initial conditions and after a sufficient amount of time the concentration of node 3 will have converged on either 1 or 0. This value is taken as the output for that simulation. We then repeat the simulation for the same pairs of inputs but this time with a normally distributed random variable added to each of them and the initial concentration of node 0 adjusted. We compare the results of this new simulation with the simulation of the same pair of input concentrations.


\section*{Analytical methods for stochastic resonance in a simple perceptron}

 In order to examine the mechanism of this effect  we will consider  a simplification of this system. The perceptron is a well-established concept and the most simple manifestation of a neural network. It  performs linear separation on $n$ inputs (see eqs. \ref{sum1} and \ref{percep1}) as described earlier but will give us a form which is easier to examine (see Fig.\ref{fig_2}a). 
This figure shows a 2-input perceptron, but the similar scheme applies for $n$-input configurations. 

\subsection*{Stochastic resonance in a simple perceptron}

Here we will spoil the threshold value $T$ (for example, by changing it from 1 to 1.3) and measure the accuracy of the output assignment functionality of the algorithm both with and without noise. The noise is represented by a random variable from a Gaussian distribution with mean $\mu = 0$ and variance $\sigma^2 = z$ where z will be increased from 0 to 1 to represent our increasing intensity of noise. Our accuracy measure simply gives the success rate with which the algorithm correctly classified all members of the test set when given the new, spoiled system to work with. In practice we calculate this using a Monte-Carlo style approach, a set of N (10000 in practice) input vectors are generated, each consisting of $n$ uniformly distributed random entries from the range $[0,1]$. We classify this set according to the unspoiled threshold $T_1$ to get 
\[
O(i) , \hspace{1cm} (i=0,1,\dots,10000)
\]  
These 10000 input vectors have a normally distributed random variable added to each element of them and are then classified again according to the spoiled threshold $T_2$ to give 
\[
D(i) , \hspace{1cm} (i=0,1,\dots,10000)
\]  
And accuracy is calculated as in \eqref{acc}. Accuracy vs noise intensity is shown in Fig.\ref{fig_2}b and again shows clearly defined SR manifesting itself in restoration of classification accuracy for an optimal amount of input noise.

\subsection*{Analytical Investigation}

Now that we have demonstrated the existence of this effect of stochastic resonance it would be prudent to attempt to explain it analytically. In doing so we must think more rigourously about our random variables and their distribution functions. For simplicity let us first consider the 2-input system with $w_1$ = $w_2$. Our input values $x_1$ and $x_2$ are uniformly distributed random variables between 0 and 1, and as such will each have a probability distribution function (pdf) given by:
\begin{equation}
f(x) = \left\{ 
  \begin{array}{l l}
    1 & \quad \text{if $x \in (0,1)$}\\
    0 & \quad \text{otherwise}\\
  \end{array} \right.
\end{equation}
But when we consider the sum of 2 uniform random variables we must adjust the pdf. Let $X = x_1 w_1 + x_2 w_2$, then
\begin{equation}
f(s) =  \left\{ 
  \begin{array}{l l}
    s & \quad \text{if $s \in (0,w)$}\\
    2w - s & \quad \text{if $s \in (w,2w)$}\\
     0 & \text{otherwise} \\
  \end{array} \right.
\end{equation}

Now we must consider the addition of noise. Our noise is distributed normally, always with $\mu = 0$ but with changing variance $\sigma ^2$. The normal distribution has the following pdf:
\begin{equation}
g(x) = \frac{1}{\sigma \sqrt{2 \pi}}e^{-\frac{x^2}{2 \sigma^2}}
\end{equation}

We wish to combine 2 lots of noise (1 for each input) and as in this case both of our weights are the same we can simply combine our noise terms together and add them to $X$. The sum of two independent normal variables is normally distributed itself with $\mu = \mu_1 + \mu_2$ and $\sigma^2 = \sigma_1^2 + \sigma_2^2$ giving us $\sigma_{new} = \sqrt{2} \sigma_{old}$.  We also wish to multiply the noise by the weight value, to do this we simply multiply the standard deviation $\sigma$ by $w$.  \\
This gives:
\begin{equation}
g(x) = \frac{1}{2 w \sigma \sqrt{\pi}}e^{-\frac{x^2}{4 w^2 \sigma^2}}
\end{equation}
Now we must consider how to formulate the system as a whole. We will consider the concept of accuracy as before, where for each point which is successfully allocated before spoiling, we consider the probability of it still being successfully allocated after spoiling and noise. Now let $T_1$ be the pre-spoiling threshold and $T_2$ be the post-spoiling threshold.
For a point $x$ this probability is given by:

If $x < T_1$
\begin{equation}
P_1(x) = \int_{-\infty}^{T_2 - x} g(y)\,dy.   
\end{equation}
And, if $x > T_1$
\begin{equation}
P_2(x) = \int_{T_2 - x}^{\infty} g(y)\,dy. 
\end{equation}

Now we must apply this to the entire distribution of $x$, remembering to use our probability density function $f(x)$. As the distribution of $x$ is independent to the distribution of the additive noise we can simply multiply the probabilities:
\begin{equation}
P_{total} = \int_0^{T_1} f(x) P_1(x)\, dx + \int_{T_1}^{2 w} f(x) P_2(x)\, dx
\end{equation}
Gives the total likelihood of 
\begin{equation}
\bigg(P\left[x<T_1\right]\wedge P\left[(x+\text{noise})<T_2\right] \bigg)+ \bigg(  P\left[x>T_1\right]\wedge P\left[(x+\text{noise})>T_2\right] \bigg),
\end{equation}
which is analogous to our measure of accuracy from before but in a more continuous sense.

This expression containing double integrals can be tidied up in terms of the error function:
\begin{equation}
\text{erf}(x) = \frac{2}{\sqrt{\pi}} \int_0^x e^{-t^2} \, dt
\end{equation}
Which is related to the cumulative distribution function of the normal distribution such that our equation now becomes:
\begin{equation}
P_{total} = \frac{1}{2} \left[ \int_0^{T_1} f(x) (1+ \text{erf}(\tfrac{T_2 - x}{2w \sigma}))\, dx + \int_{T_1}^{2 w} f(x) (1- \text{erf}(\tfrac{T_2 - x}{2w \sigma}))\, dx \right]
\end{equation}
As $f(x)$ represents a probability distribution function we know that its integral over its whole range ($[0,2w]$ in this case) must be equal to 1. Using this we can simplify further:
\begin{equation}
P_{total} = \frac{1}{2} \left[ 1 +  \int_0^{T_1} f(x) \text{erf}(\tfrac{T_2 - x}{2w \sigma})\, dx - \int_{T_1}^{2 w} f(x)  \text{erf}(\tfrac{T_2 - x}{2w \sigma})\, dx \right]
\end{equation}

\begin{figure}[!h]
\centering\includegraphics[width=2.5in]{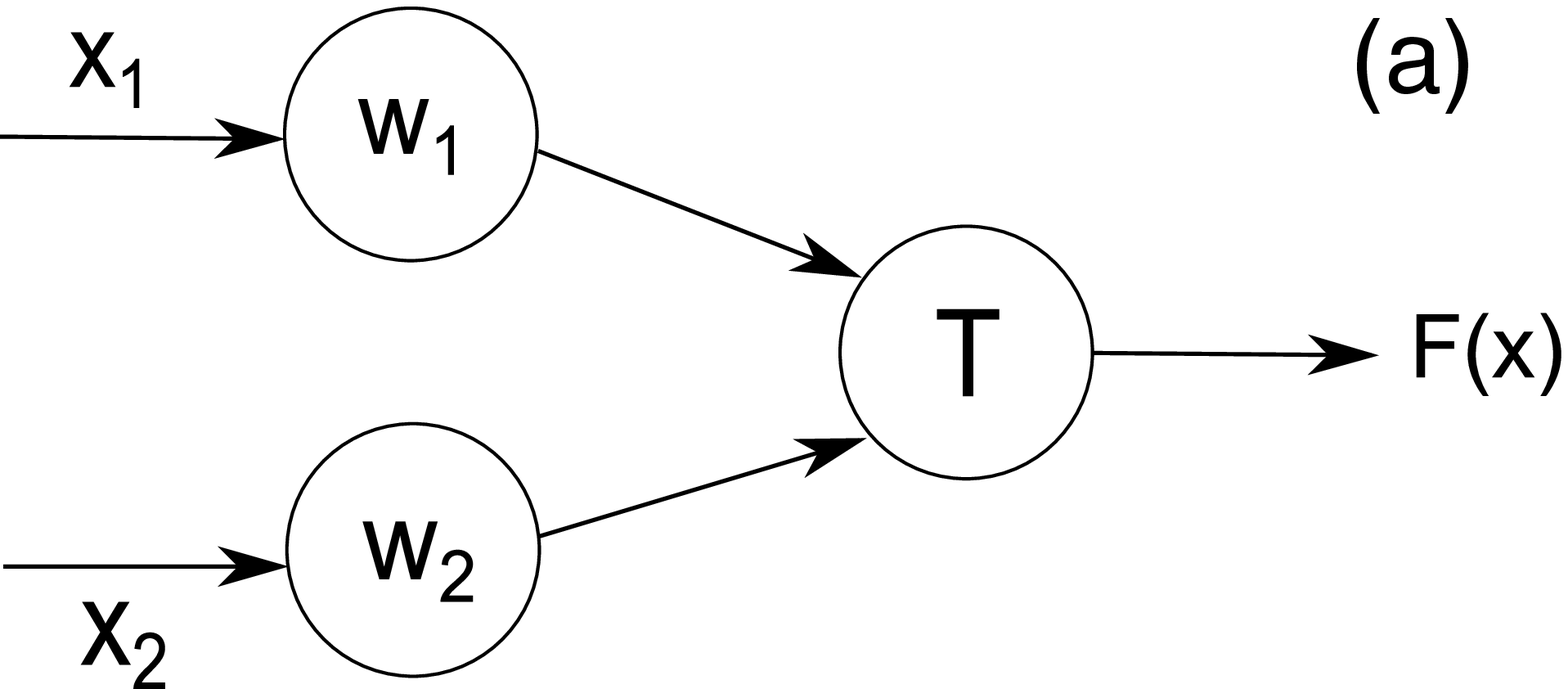}
\hspace{0.5cm}\centering\includegraphics[width=2.5in]{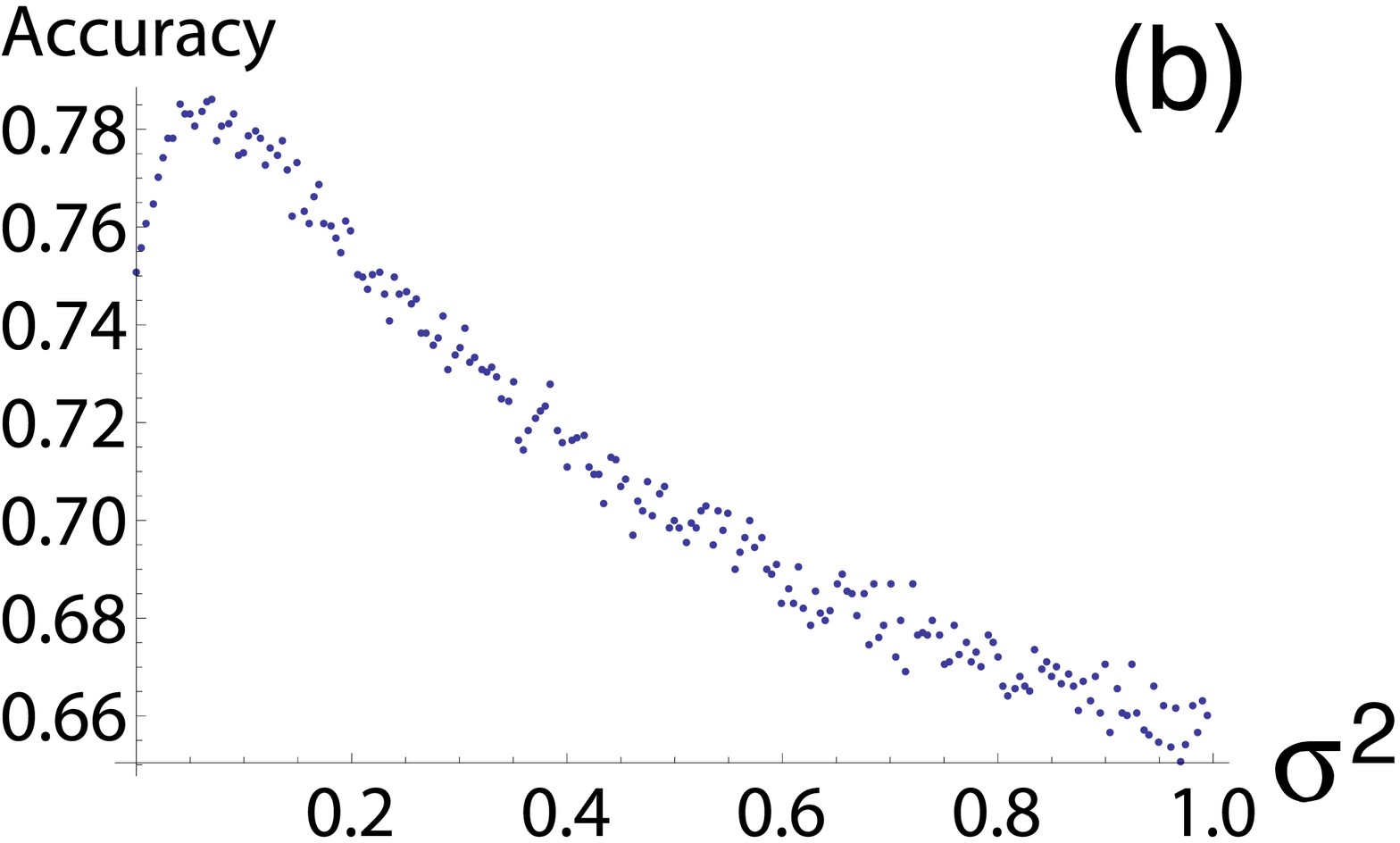}

\vspace{0.5cm}
\centering\includegraphics[width=2.5in]{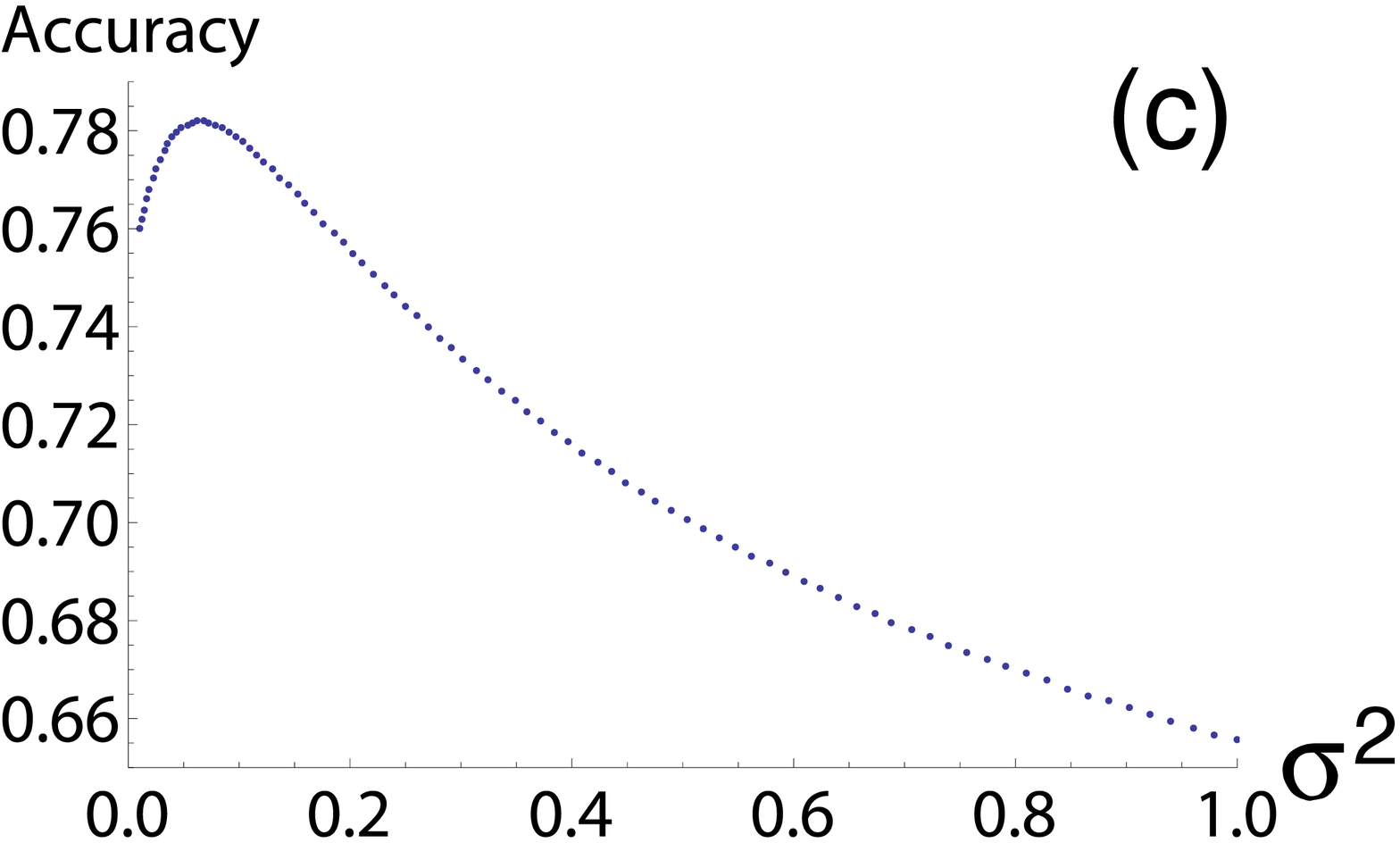}
\hspace{0.5cm}\centering\includegraphics[width=2.5in]{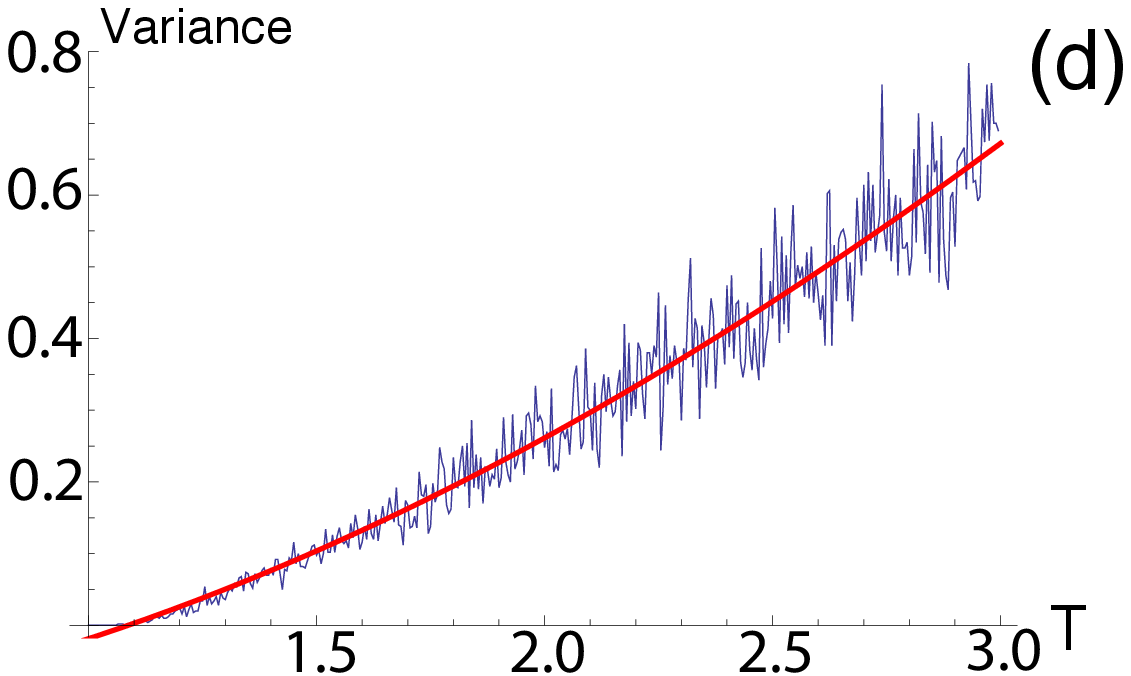}
\caption{(a) Visualisation of a simple perceptron with two inputs $x_1$ and $x_2$. (b) Graph showing accuracy vs. noise intensity in a threshold spoiled from 1 to 1.3. This is a well known bell shaped curve for SR. (c) Analytical results displayed here closely matches numerical simulations in (b). (d) Linear correlation between optimal noise intensity and spoiled threshold value.}
\label{fig_2}
\end{figure}

By plotting this for increasing variance $\sigma^2$ and comparing it with the simulated results using the same coefficients we can see perfect correlation (compare Fig.\ref{fig_2}b and c). From this one can also find optimal noise intensity as a function of a spoiled threshold value (Fig. \ref{fig_2}d).

\subsection*{Generalising the Model}
To  generalise this model we must consider a system with $n$ inputs and independent weights for each input $(w_1,w_2,\dots,w_n)$. First consider our input variable 
$X \sim w_1 X_1 + w_2 X_2 + \dots + w_n X_n$.
Each $w_i X_i$ is a uniformly distributed random variable on the range $(0,w_i)$. We can generate the probability distribution function for a distribution such as this from the following function ~\cite{sadooghi-alvandi_distribution_2007}.
\begin{equation}
f_n(s) = \frac{1}{W_n (n-1)!} \left\{  s^{n-1}  + \sum_{k=1}^{n} (-1)^k \sum_{J_k} \left[ \left( s - \sum_{l=1}^{k} w_{j_l} \right) _+ \right] ^{n+1} \right\}
\end{equation}

Where:
 \[
J_k = \{(j_1,j_2,\dots,j_k) : 1 \leq j_1 < j_2 < \dots< j_k \leq n\}
\]
 \[
x_+ = \text{max}(x, 0)
\]
\[
W_n = \prod_{k=1}^n w_n
\]

Theoretically it is now possible to produce the probability distribution function for any number of independent inputs but due to the nature of this equation it is difficult to treat this general formula analytically. We will proceed by generating it for the case $n=2$.

If we evaluate this for $n = 2$ we arrive at the probability distribution function for 2 independently weighted inputs. Labelling such that $w_1 < w_2$
\begin{equation}
f_2(s) =  \left\{ 
  \begin{array}{l l}
    \frac{s}{w_1 w_2} & \quad \text{if $s \in (0,w_1)$}\\
    \frac{1}{w_2} & \quad \text{if $s \in (w_1,w_2)$}\\
   \frac{1}{w_1 w_2}(w_1 + w_2 -s) & \quad \text{if $s \in (w_2,w_1 + w_2)$}\\
     0 & \text{otherwise} \\
  \end{array} \right.
\end{equation}

Considering the addition of $n$ independently weighted noise terms is simpler. We can simply consider a single Gaussian distribution with a modified variance $\sigma _{new}^2$ due to various properties of adding normal distributions and multiplying them by constants:

Let
\[
 X \sim \mathcal{N}(0,\sigma^2)
\]
Then
\[
a X \sim \mathcal{N}(0,a^2 \sigma^2)
\]
And
\[
aX_1 + b X_2 \sim \mathcal{N}(0,a^2 \sigma _1 ^2 + b^2 \sigma_2 ^2)
\]

From this we can see how to create our new variance. 
\[
\sigma _{new} ^2 = \sum_{i = 1}^n w_i ^2 \sigma_i ^2
\]
But all $\sigma _i$ will be equal so we have:
\[
\sigma _{new} ^2 = \sigma_{old} ^2 \sum_{i = 1}^n w_i ^2  
\]
We have now reduced our $n$ input system to a single variable distributed according to $f_n(s)$ and a single noise term, defined as above. 

Defining $\sigma$ as stated and defining $W$ as the sum of all weights we end up with a similar description for the full system as before.
\begin{equation}
\label{final}
P_{total}(\sigma) = \frac{1}{2} \left[ 1 +  \int_0^{T_1} f_n(x) \text{erf}(\tfrac{T_2 - x}{\sqrt{2}\sigma_{new}})\, dx - \int_{T_1}^{W} f_n(x)  \text{erf}(\tfrac{T_2 - x}{\sqrt{2}\sigma_{new}})\, dx \right]
\end{equation}
This fully describe the accuracy vs. noise intensity curve for $n$ inputs with independent weightings.

\subsection*{Optimal Spoiling}

Earlier we discussed the idea of threshold spoiling acting as some sort of mechanism for providing robustness to noise. Now with this in mind we can refer back to \eqref{final} but rather than considering it as a function of $\sigma$ we take $T_2$ to be the independent variable and $\sigma$ to be a parameter. 

\begin{equation}
\label{finalt2}
P(T_2;\sigma) = \frac{1}{2} \left[ 1 +  \int_0^{T_1} f_n(x) \text{erf}(\tfrac{T_2 - x}{\sqrt{2}\sigma_{new}})\, dx - \int_{T_1}^{W} f_n(x)  \text{erf}(\tfrac{T_2 - x}{\sqrt{2}\sigma_{new}})\, dx \right]
\end{equation}

\begin{figure}[!h]
\centering\includegraphics[width=2.5in]{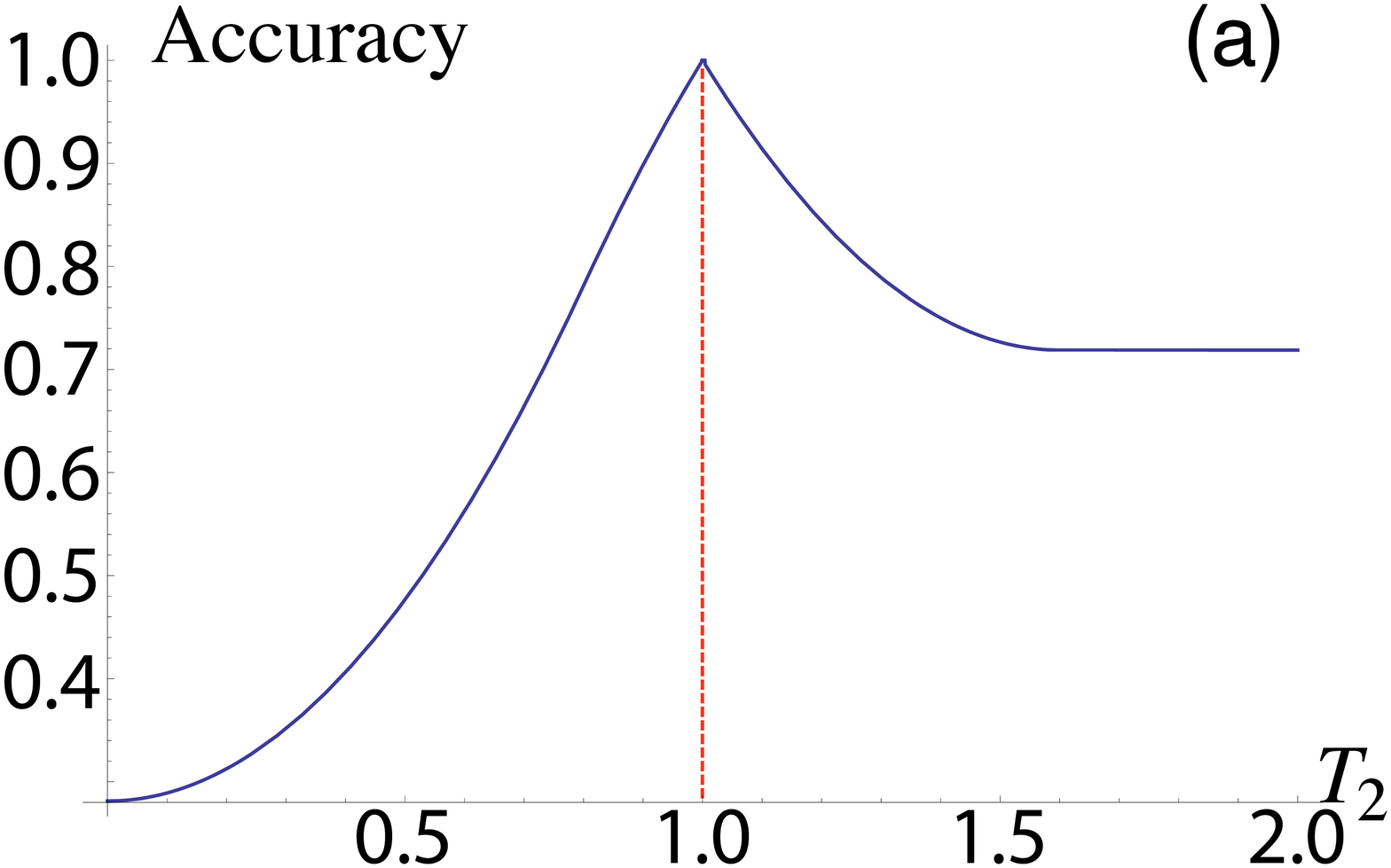}
\hspace{0.2cm}\centering\includegraphics[width=2.5in]{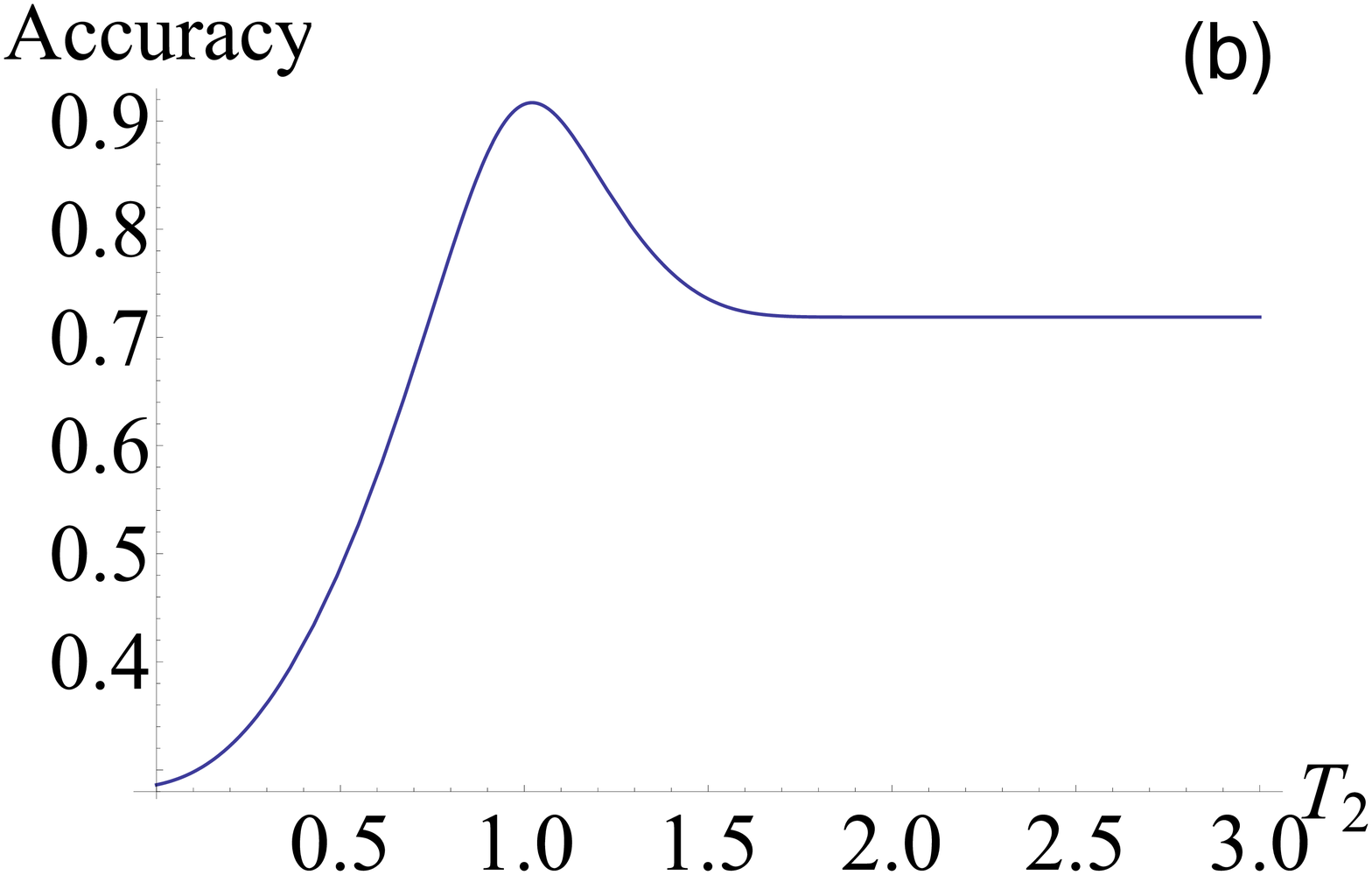}

\vspace{0.5cm}\centering\includegraphics[width=2.5in]{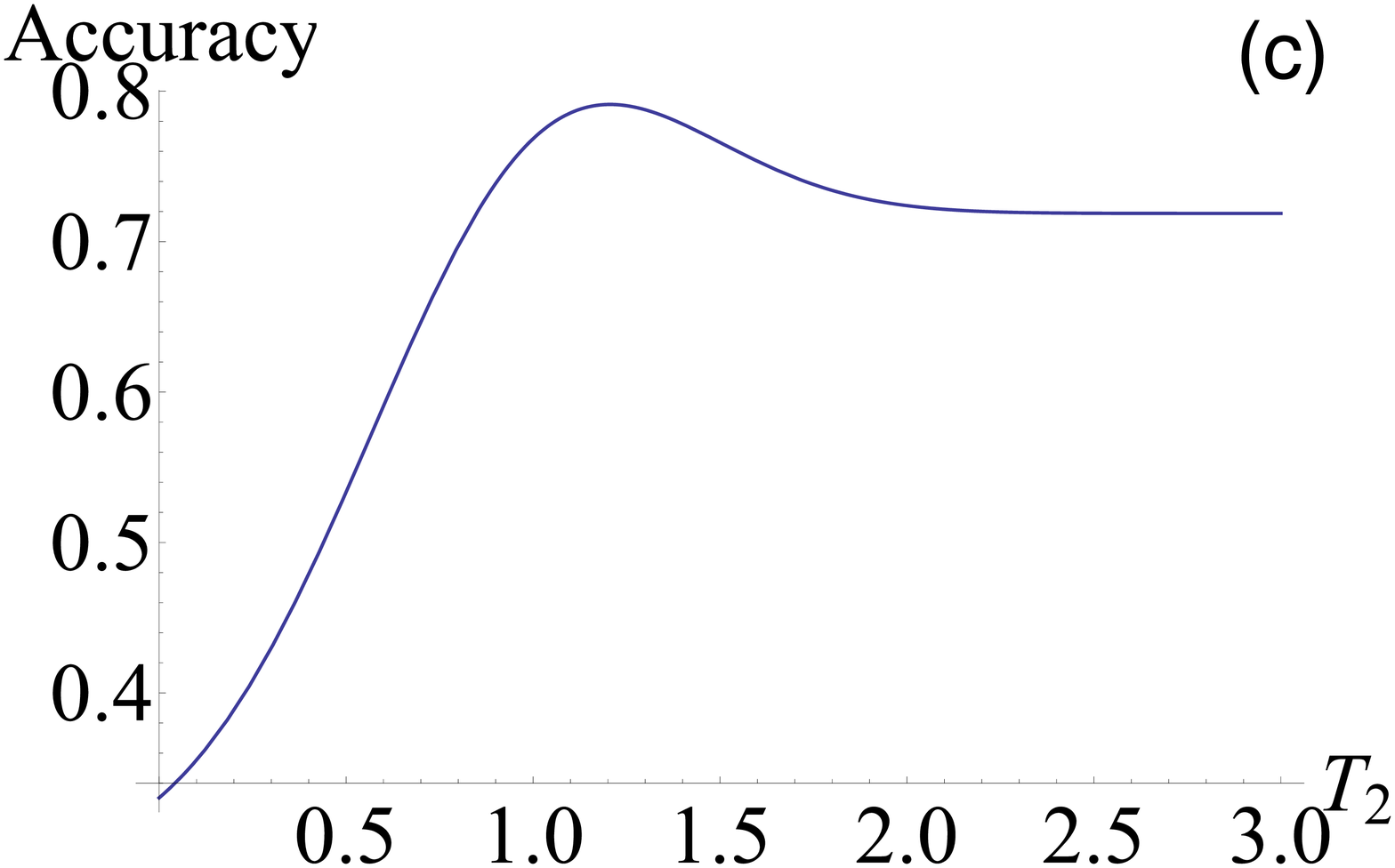}
\hspace{0.2cm}\centering\includegraphics[width=2.5in]{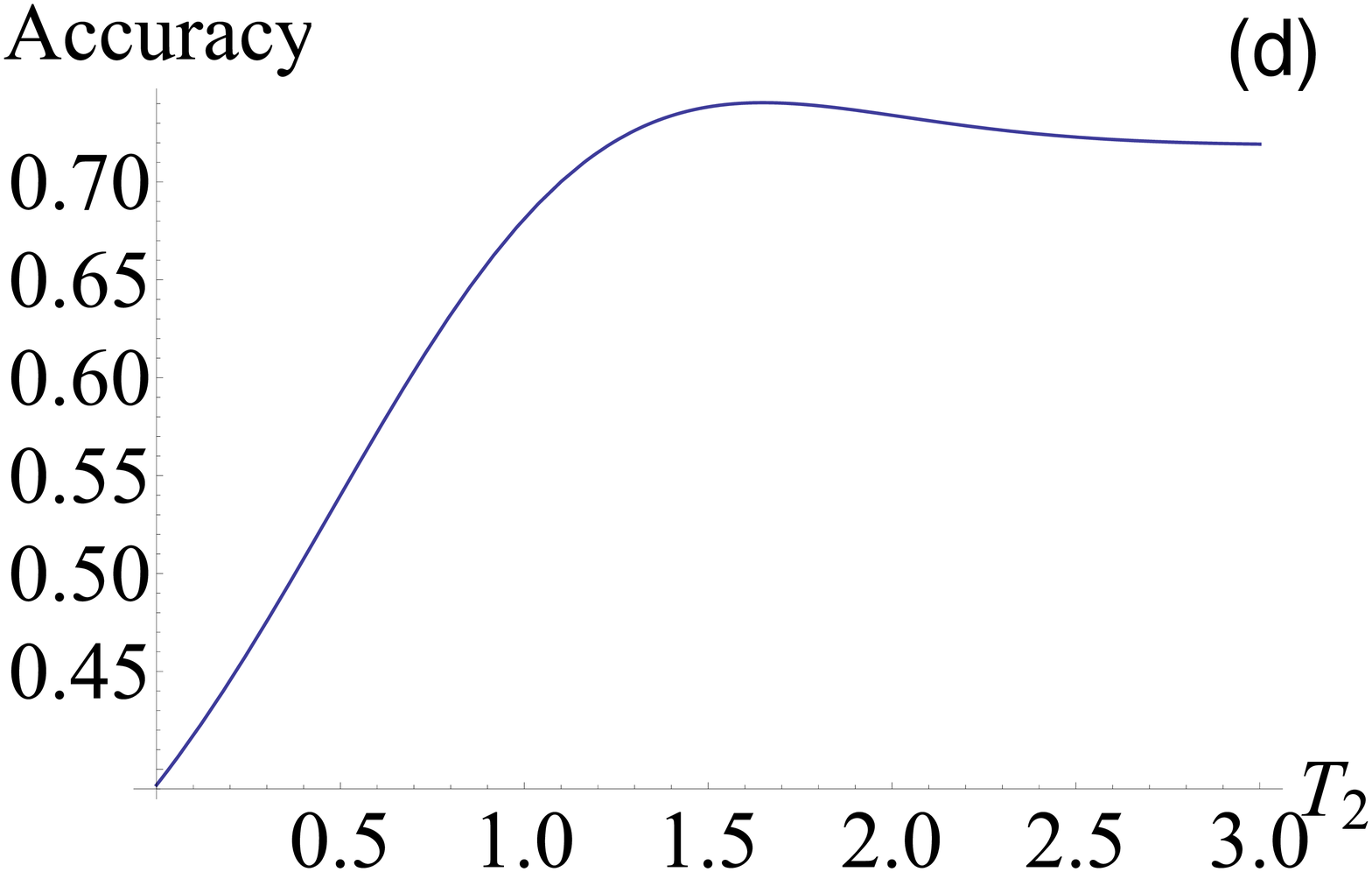}
\caption{(a) Here the function \eqref{finalt2} is plotted for $\sigma^2=0$. Examples of increasing values for $\sigma^2$ showing an increasing value for optimal $T_2$, this corroborates the findings of fig.\protect\ref{fig_2}d which showed a linear relationship between spoiling and optimal noise. Increase of noise intensity leads to the shift of the optimal threshold to the right, see (b-d). Here b: $\sigma^2=0.01$, c: $0.1$, and d: $0.2$.
}
\label{fig_3}
\end{figure}

Plotting this for a given value of noise intensity will show us the optimal spoiling value to achieve the greatest value of accuracy. As shown in Fig.\ref{fig_3}a we can confirm that for a non-noisy system the optimal value for $T_2$ is $1$ and equal to  $T_1$ as would be expected. Increasing the noise intensity also increases the optimal $T_2$ (Fig.\ref{fig_3} b-d).

If the noise is only intrinsic, re-orienting our equation in this way in fact makes more sense as the threshold value in our original model would be much more flexible than the intrinsic intensity of noise in the system. Thermodynamic considerations tell us that to expect an increase in noise in a chemical system we would require either a decrease in cell volume or an increase in temperature. In fact studies have shown~\cite{partridge_evolution_1994} that an increase in environmental temperature will cause an evolutionary increase in cell volume in \emph{Drosophila melanogaster}. Aside from this it seems unlikely that optimising behaviours such as those we are considering would provide significant enough evolutionary pressure on cell size for it to be sensible to consider it a variable in this sense. 

In a potential synthetic biology implementation of such a linear perceptron \eqref{finalt2} could be invaluable in terms of optimizing the system. It allows us to determine how best to skew the threshold parameter in order to compensate for the inescapable effects of noise in the system. Indeed this could be applied to any application of linear classification attempting to operate in a noisy environment.

In addition to simply a practical application it is of interest to try and explore such systems as designs of this type appear frequently in nature. For example, it has been shown that the process by which bristles in the epithelial cells of \emph{Drosophilia} are organised could be imagined as a perceptron style system~\cite{cohen_dynamic_2010}. Delta-Notch signalling allows cells to make analysis from numerous inputting filopodia. From this information the cell can make a decision as to whether it should differentiate to a bristle producing hair.

\section*{Summary}

In summary we have outlined a design by which linear separation can be performed genetically on the concentrations of certain input protein concentrations. In addition to this we are able to contextualize this separation through the introduction of external kinases which allow the systems weightings (and thus its line of separation) to be adjusted within the cellular life rather than having to be adjusted evolutionarily. Considering this system in a presence of noise  we observed, demonstrated and quantified the effect of stochastic resonance in linear classification systems with input noise and threshold spoiling. To explain a mechanism of this effect we have considered a simple classifying perceptron and have shown that analytical results match the numerical simulations. The consideration of linear separation in noisy environments is relevant due to the inherantly noisy nature of gene expression in cells both due to extrinsic or intrinsic factors. 

\section*{Discussion}

\subsection*{Spoiled vs not spoiled perceptron in the presence of noise}

\begin{figure}[!h]
\centering\includegraphics[width=2.8in]{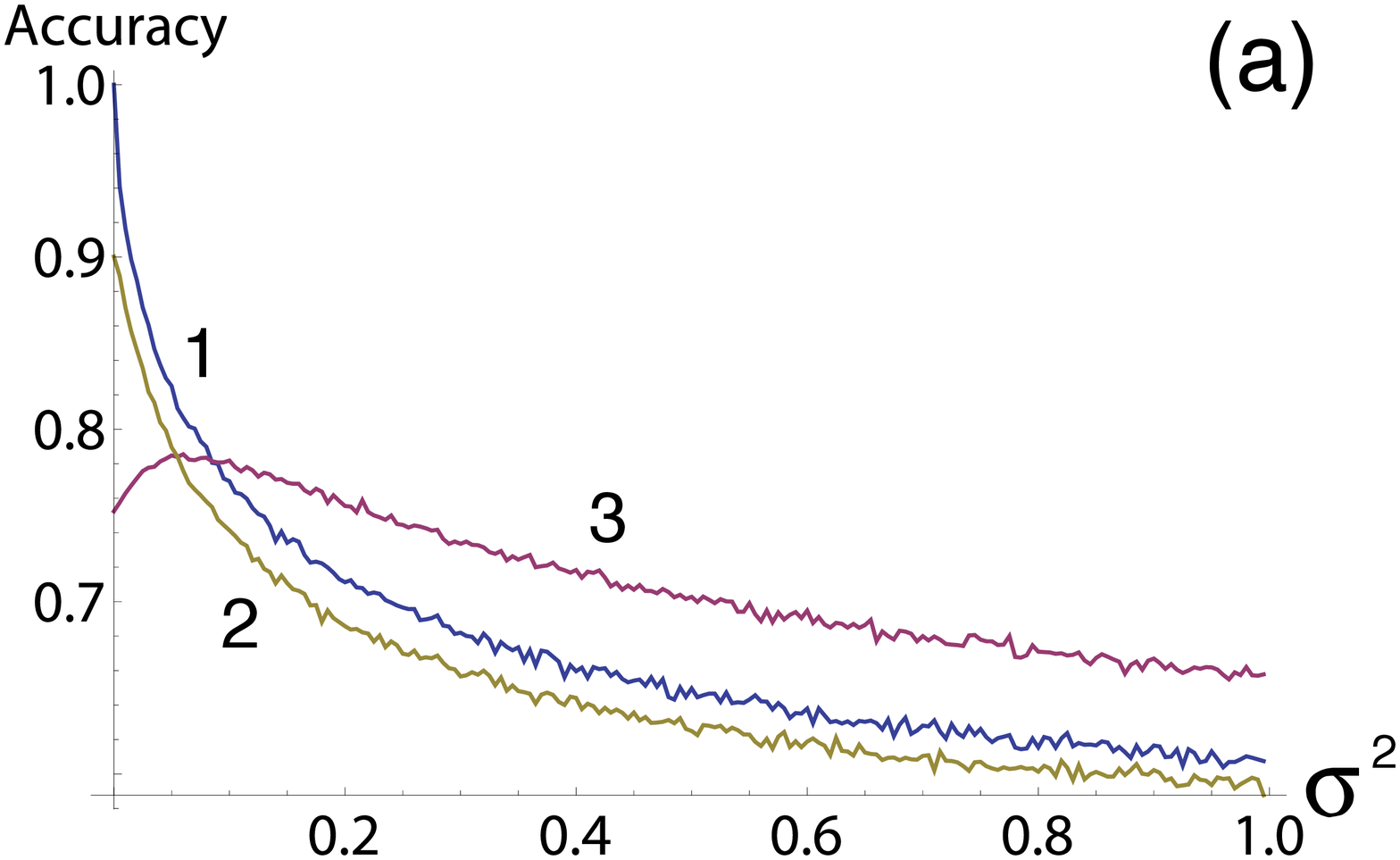}
\centering\includegraphics[width=2.4in]{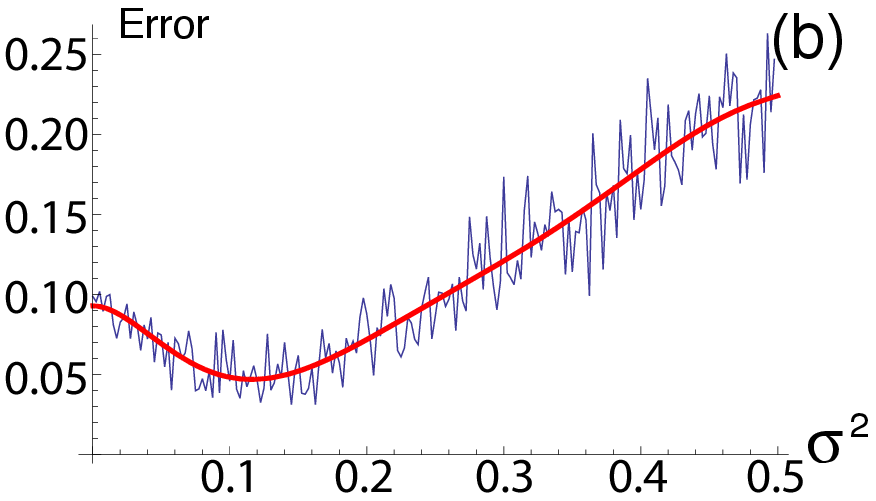}
\caption{(a) Accuracy vs. noise intensity for various values of threshold spoiling. Here we see 3 curves for various values of threshold spoiling. The blue line shows the case for zero threshold spoiling, here we see perfect accuracy for no noise, as would be expected but a sharp decrease with the addition of noise. The purple line shows a threshold spoiling which exhibited some degree of stochastic resonance, not only do we see a peak at which it exceeds the unspoiled accuracy but we also see it far exceeding the performance of the unspoiled threshold throughout the intensity range examined. Essentially the threshold spoiling has provided some degree of robustness to the noise which is an extremely interesting property in itself. The contrary could also be inferred that the noise is providing some degree of robustness to threshold spoiling by a `blurring' of the lines as seen in Fig.\protect\ref{fig_1}b.
This figure clearly demonstrates that there is a strong relationship between threshold spoiling, input noise and accuracy but how they work together can be highly variable. (b) Graph showing learning resonance for spoiled T=1.1, $w_1=0.7$ $w_2=0.7$.  }
\label{fig_4}
\end{figure}

Whereas an idea to spoil the threshold of classification may look a bit artificially in a sense that we first corrupt a classification, and then restore it, the Fig.\ref{fig_4}a
illustrates that for certain noise intensities spoiling is a genuine way how a biological system could adapt to the unavoidable level of stochasticity. This figure shows that for noise intensity $\sigma^2>0.1$ spoiled perceptron has better accuracy in classification than not spoiled one. It could lead to speculations that, since some noise is
certainly present in gene expression, classifying genetic networks will evolve towards shifting the classification threshold to compensate an effect of noise.

\subsection*{Resonance in the Learning Algorithm}

The design presented allows for the `contextualisation' of classification whereby the same network implemented in a different scenario could perform an entirely different separation task depending on the stimuli present. While this allows for a great deal of flexibility we would like to explore the idea of the implementation of some kind of perceptron learning in the same vein as the delta learning rule, a simple version of back propagation learning. In this learning method weights are updated by a process of comparison of the perceptron output to some ideal result for the given inputs. Mathematically we consider it as follow. Let $\textbf{w} = \left\{ w_1, \dots ,w_n \right\} $ denote the desired weight set and $\textbf{$\bar{w}^l$} =  \left\{w^l_1, \dots w^l_n \right\} $ denote the learning algorithms $l$th iterative attempt at learning the desired weights. Also let $\chi = \bigg\{ X^j : x^j = \left\{x_1^j, \dots, x_n^j \right\} \bigg\}$ be the learning set. Each $x^j$ is a set of input values for which we have the desired output given by $O(x^j)$. Then we update our weights as follows,
\[
w_i^{l+1} = w_i^l + \alpha(O(x^l)-D(x^l))x_i^l    
\]
Where
\[
D(x^l) = \left\{ 
  \begin{array}{l l}
    1 & \quad \text{if $\sum_i x_i w_i^l > T$}\\
    0 & \quad \text{if $\sum_i x_i w_i^l  \leq T$}\\
  \end{array} \right.
\]

While this is more computationally intensive than simply setting our weights as desired as in the current implementation it has a major advantage in the sense that it is results driven. With this kind of learning we are guarenteed for our system to conform to our desired outputs as provided in the test sets. When setting weights directly we are not guarenteed that we will get the results we desire, simply that we will have the weights given. 

The thinking behind backpropagation learning involves the system performing a comparison between the result which it has arrived at and some ideal result. By considering discrepancies between these two results it can then make adjustments to the process by which it arrived at its result in order to move closer to the ideal result. 

This kind of learning requires a large degree of reflexivity on the behalf of the network and as such it is difficult to imagine a biological implementation of such a learning technique. Despite this we feel it is worth considering whilst on the topic of perceptrons as it is an extremely effective technique. The procedure is simple: for each element $j$ of some training set O, where we have a set of inputs along with their desired output, we should compute the following:
\[
X_j = \sum_{i=1}^{n} x_i. w_i
\]
\[
F(X_j) = \left\{ 
  \begin{array}{l l}
    1 & \quad \text{if $X_j > T$}\\
    0 & \quad \text{if $ X_j \leq T$}\\
  \end{array} \right.
\]

\[
\delta(j) = O_j - F(X_j)
\]
\[ 
w_{i,new} = w_{i,old} + \alpha \delta(j) x_i 
\]

Here $\alpha$ is some sort of learning rate, if we choose $\alpha$ too large then we may end overcorrect and adjusting the weights too far in each update, but the smaller it is chosen then the longer it will take to arrive at the desired weights.

This improvement to $w_i$ is performed for each $i$ and looped through for each entry in the training set. The whole process is then repeated until the error in the system is sufficiently small or a maximum number of iterations is reached. For a training set of $N$ input/outputs we have:
\[ error = \frac{\sum_{j=1}^N \delta(j)}{N}
\]
For an error threshold of 0.001, $\alpha = 0.01$ was found to be suitable for optimal learning speeds and for what follows, the unspoiled threshold has $T=1$.

Previously, we have just been examining the output assignment functionality of the algorithm but the ideas of stochastic resonance could also be applied to the learning algorithm. As in the previous case we must spoil the system in some way and then we will examine some kind of measure of accuracy for increasing intensity of noise. 
The way in which this will be implemented is by spoiling the value of $T$ in $F(X)$ as defined in the previous section and adding the noise to each $x_1$ and $x_2$ from the test set before it is fed into $F(x)$.
The accuracy is measured by considering the error from the actual intended weights. This is done by finding the Euclidean distance of the algorithms attempts at the weights to the correct values:
\[
	error = \sqrt{(w_1 - w_{1,new})^2 + (w_2 - w_{2,new})^2 }
\]

The algorithm was allowed to run through the test set 2000 times before arriving at its final attempt at the correct weights $w_{1,new}$ and $w_{2,new}$.
As is clear from Fig.\ref{fig_4}b we again find that the system performs optimally under some non-zero amount of noise, shown here by a minimized error (where noise intensity $\sigma^2 \approx 0.12$) rather than a maximized accuracy rate as before.

Another simulation featuring independently changed weights shows that the resonance still applies in this case.
It would be of interest to perform a more thorough investigation into this effect and to try and describe it analytically if possible. A better understanding of this effect and the circumstances in which we can find it would be of great interest for applications in learning algorithms in general.
Of additional interest would be to investigate the possibility of a genetic implementation of learning of this kind. It is not clear whether such an implementation would be possible, if it were possible it would undoubtably be much more complex than our linear classification network. If possible it's construction and simulation would certainly be of great interest.




Finally, construction of intelligent intracellular gene regulating networks is the hot topic of synthetic biology, e.g., see for a review \cite{2009_Lu}, and here we have shown that unavoidable noise can be constructively used in such design.

\section*{Acknowledgment}

We acknowledge support from CR-UK and Eve Appeal funded project PROMISE-2016, the Deanship of Scientific Research (DSR), King Abdulaziz University (KAU), Jeddah,  Russian Foundation for Basic Research (14-02-01202), and partly by the grant (the agreement of
August 27, 2013 N 02.÷.49.21.0003 between The Ministry of education and
science of the Russian Federation and Lobachevsky State University of Nizhni Novgorod).


\begin{thebibliography}{10}
\providecommand{\url}[1]{\texttt{#1}}
\providecommand{\urlprefix}{URL }
\expandafter\ifx\csname urlstyle\endcsname\relax
  \providecommand{\doi}[1]{doi:\discretionary{}{}{}#1}\else
  \providecommand{\doi}{doi:\discretionary{}{}{}\begingroup
  \urlstyle{rm}\Url}\fi
\providecommand{\bibAnnoteFile}[1]{%
  \IfFileExists{#1}{\begin{quotation}\noindent\textsc{Key:} #1\\
  \textsc{Annotation:}\ \input{#1}\end{quotation}}{}}
\providecommand{\bibAnnote}[2]{%
  \begin{quotation}\noindent\textsc{Key:} #1\\
  \textsc{Annotation:}\ #2\end{quotation}}
\providecommand{\eprint}[2][]{\url{#2}}

\bibitem{saigusa_amoebae_2008}
Saigusa T, Tero A, Nakagaki T, Kuramoto Y (2008) Amoebae anticipate periodic
  events.
\newblock Physical Review Letters 100.
\input{saigusa_amoebae_2008}

\bibitem{1991_Hjelmfelt}
Hjelmfelt A, Weinberger ED, Ross J (1991) Chemical implementation of neural
  networks and turing machines.
\newblock Proc Natl Acad Sci U S A 88: 10983-7.
\input{1991_Hjelmfelt}

\bibitem{hjelmfelt_implementation_1995}
Hjelmfelt A, Ross J (1995) Implementation of logic functions and computations
  by chemical kinetics.
\newblock Physica D: Nonlinear Phenomena 84: 180--193.
\input{hjelmfelt_implementation_1995}

\bibitem{1994_Bray}
Bray D, Lay S (1994) Computer simulated evolution of a network of
  cell-signaling molecules.
\newblock Biophys J 66: 972-7.
\input{1994_Bray}

\bibitem{2007_Gandhi}
Gandhi N, Ashkenasy G, Tannenbaum E (2007) Associative learning in biochemical
  networks.
\newblock J Theor Biol 249: 58-66.
\input{2007_Gandhi}

\bibitem{fernando_molecular_2008}
Fernando CT, Liekens AM, Bingle LE, Beck C, Lenser T, et~al. (2008) Molecular
  circuits for associative learning in single-celled organisms.
\newblock Journal of The Royal Society Interface 6: 463--469.
\input{fernando_molecular_2008}

\bibitem{2007_Jones}
Jones B, Stekel D, Rowe JE, Fernando C (2007) Is there a liquid state machine
  in the bacterium escherichia coli?
\newblock In: Proceedings of IEEE Symposium on Artificial Life : 187-191.
\input{2007_Jones}

\bibitem{2011_Qian}
Qian L, Winfree E, Bruck J (2011) Neural network computation with dna strand
  displacement cascades.
\newblock Nature 475: 368-72.
\input{2011_Qian}

\bibitem{mcadams_its_1999}
{McAdams} HH, Arkin A (1999) It's a noisy business! genetic regulation at the
  nanomolar scale.
\newblock Trends in Genetics 15: 65--69.
\input{mcadams_its_1999}

\bibitem{elowitz_stochastic_2002}
Elowitz MB (2002) Stochastic gene expression in a single cell.
\newblock Science 297: 1183--1186.
\input{elowitz_stochastic_2002}

\bibitem{gillespie_chemical_2000}
Gillespie DT (2000) The chemical langevin equation.
\newblock The Journal of Chemical Physics 113: 297.
\input{gillespie_chemical_2000}

\bibitem{1981_Benzi_jpa}
Benzi R, Sutera A, Vulpiani A (1981) The mechanism of stochastic resonance.
\newblock J Phys A 14: L453.
\input{1981_Benzi_jpa}

\bibitem{russell_use_1999}
Russell DF, Wilkens LA, Moss F (1999) Use of behavioural stochastic resonance
  by paddle fish for feeding.
\newblock Nature 402: 291--294.
\input{russell_use_1999}

\bibitem{priplata_vibrating_2003}
Priplata AA, Niemi JB, Harry JD, Lipsitz LA, Collins JJ (2003) Vibrating
  insoles and balance control in elderly people.
\newblock The Lancet 362: 1123--1124.
\input{priplata_vibrating_2003}

\bibitem{2002_Mori_prl}
Mori T, Kai S (2002) Noise-induced entrainment and stochastic resonance in
  human brain waves.
\newblock Physical Review Letters 88: 218101.
\input{2002_Mori_prl}

\bibitem{UsherF00}
Usher M, Feingold M (2000) Stochastic resonance in the speed of memory
  retrieval.
\newblock Biological Cybernetics 83.
\input{UsherF00}

\bibitem{alberts_molecular_2008}
Alberts B (2008) Molecular biology of the cell.
\newblock New York: Garland Science.
\input{alberts_molecular_2008}

\bibitem{suzuki_oscillatory_2011}
Suzuki N, Furusawa C, Kaneko K (2011) Oscillatory protein expression dynamics
  endows stem cells with robust differentiation potential.
\newblock {PLoS} {ONE} 6: e27232.
\input{suzuki_oscillatory_2011}

\bibitem{1997_Weiss}
Weiss JN (1997) The hill equation revisited: uses and misuses.
\newblock FASEB J 11: 835-41.
\input{1997_Weiss}

\bibitem{sadooghi-alvandi_distribution_2007}
Sadooghi-Alvandi SM, Nematollahi AR, Habibi R (2007) On the distribution of the
  sum of independent uniform random variables.
\newblock Statistical Papers 50: 171--175.
\input{sadooghi-alvandi_distribution_2007}

\bibitem{partridge_evolution_1994}
Partridge L, Barrie B, Fowler K, French V (1994) Evolution and development of
  body size and cell size in drosophila melanogaster in response to
  temperature.
\newblock Evolution 48: pp. 1269-1276.
\input{partridge_evolution_1994}

\bibitem{cohen_dynamic_2010}
Cohen M, Georgiou M, Stevenson NL, Miodownik M, Baum B (2010) Dynamic filopodia
  transmit intermittent delta-notch signaling to drive pattern refinement
  during lateral inhibition.
\newblock Developmental Cell 19: 78--89.
\input{cohen_dynamic_2010}

\bibitem{2009_Lu}
Lu TK, Khalil AS, Collins JJ (2009) Next-generation synthetic gene networks.
\newblock Nat Biotechnol 27: 1139-50.
\input{2009_Lu}

\end{thebibliography}

{\bf Figure legends}

\end{document}